\documentclass[manuscript]{acmart}

\usepackage[utf8]{inputenc}
\usepackage{etoolbox}

\usepackage{multicol}
\usepackage{graphicx}
\usepackage{newunicodechar}
\newunicodechar{َ}{}

\usepackage{subfigure}
\usepackage{verbatim}
\usepackage{multirow}
\usepackage{makecell}
\usepackage{changepage}
\usepackage{url}
\usepackage{comment}
\usepackage[group-separator={,}]{siunitx}
\usepackage{colortbl}
\usepackage{enumitem}
\usepackage{footnote}
\usepackage{array}
\usepackage{ragged2e}
\usepackage{xcolor}
\makesavenoteenv{tabular}
\makesavenoteenv{table}

\definecolor{light-gray}{gray}{0.95}
\definecolor{burgundy}{RGB}{144,0,32}

\usepackage{dirtytalk}
\usepackage{xparse}
\NewDocumentCommand{\quoting}{O{P}m}{``\emph{#2}''\emph{[\textnormal{#1}]}}

\newcolumntype{L}[1]{>{\RaggedRight\arraybackslash}p{#1}}

\AtBeginDocument{%
  \providecommand\BibTeX{{%
    \normalfont B\kern-0.5em{\scshape i\kern-0.25em b}\kern-0.8em\TeX}}}

\copyrightyear{2026}
\acmYear{2026}
\acmConference[CHIWORK '26]{Proceedings of the 5th Annual Symposium on Human-Computer Interaction for Work}{June 22--25, 2026}{Linz, Austria}

\setcopyright{none}

\begin{document}

\title[Vibe Coding in Product Teams]{Vibe Coding in Product Teams: Reconfiguring AI-Assisted Workflows, Prototyping, and Collaboration}





\begin{abstract}

Generative AI is reshaping product design practices through ``vibe coding'', where product team members express intent in natural language and AI translates it into functional prototypes and code. Despite rapid adoption, little research has examined how vibe coding reconfigures product development workflows and collaboration. Drawing on interviews with 22 product team members across enterprises, startups, and academia, we show how vibe coding follows a four-stage workflow of ideation, generation, debugging, and review. This accelerates iteration, supports creativity, and lowers participation barriers. However, participants reported challenges of code unreliability, integration, and AI over-reliance. We find tensions between efficiency-driven prototyping (“intending the right design”) and reflection (“designing the right intention”), introducing new asymmetries in trust, responsibility, and social stigma within teams. Through a responsible human-AI collaboration lens for AI-Assisted product design and development, we contribute a deeper understanding of deskilling, ownership and disclosure, and creativity safeguarding in the age of vibe coding.

\end{abstract}

\ccsdesc[500]{Human-centered computing~Empirical studies in HCI}

\keywords{Vibe coding, generative AI, product design, product teams, UX design, human-AI collaboration, AI-assisted programming}

\author{Jie Li}
\authornote{Equal contributions.}
\orcid{0000-0002-6791-104X}
\affiliation{
  \institution{Cake Researcher}
  \city{Delft}
  \country{Netherlands}}
   \affiliation{%
  \institution{MIT Media Lab}
   \city{Cambridge}
  \country{United States}}
\email{jasminejue@gmail.com}

\author{Youyang Hou}
\orcid{0000-0002-6453-1260}
\affiliation{
  \institution{Notion Labs}
  \city{San Francisco}
  \country{United States}}
\email{youyang@makenotion.com}

\author{Laura Lin}
\orcid{0000-0001-7550-5095}
\affiliation{
  \institution{Collov AI}
  \city{Redwood City}
  \country{United States}}
\email{lauralin.uxr@gmail.com}

\author{Ruihao Zhu}
\orcid{0009-0004-2600-5165}
\affiliation{
  \institution{Cornell University}
  \city{New York}
  \country{United States}}
\email{ruihao.zhu@cornell.edu}

\author{Hancheng Cao}
\orcid{0000-0001-7231-1076}
\affiliation{
  \department{Goizueta Business School}
  \institution{Goizueta Business School, Emory University}
  \city{Atlanta}
  \country{United States}}
\email{hancheng.cao@emory.edu}

\author{Abdallah El Ali}
\authornotemark[1]
\orcid{0000-0002-9954-4088}
\affiliation{%
  \institution{Centrum Wiskunde \& Informatica}
    \city{Amsterdam}
    \country{The Netherlands}}
  \affiliation{%
  \institution{Utrecht University}
   \city{Utrecht}
  \country{The Netherlands}
}
\email{aea@cwi.nl}
\maketitle

\section{Introduction} 
\label{sec:intro}

The recent proliferation of Generative AI (GenAI) tools has transformed how people write \cite{Liang2025widespread} and design digital experiences \cite{Takaffoli2024design}, including how product team members prototype, develop, and deploy software. Built on large language and multimodal foundation models~\cite{Bommasani2021FoundationModels}, systems such as ChatGPT\footnote{\url{https://chat.openai.com}}, GitHub Copilot\footnote{\url{https://github.com/features/copilot}}, Claude Code\footnote{\url{https://code.claude.com/}}, Cursor\footnote{\url{https://www.cursor.com}}, Lovable\footnote{\url{https://lovable.dev}}, and Bolt\footnote{\url{https://bolt.new}} translate natural-language prompts directly into working code, user interfaces, and interactive prototypes. Earlier AI-assisted tools, such as code autocompletion or small code snippet suggestions, were designed mainly to speed up single tasks without changing overall workflows~\cite{svyatkovskiy2020intellicode}. By contrast, these new vibe coding systems act as continuous partners: they can generate entire components, integrate APIs, and adapt outputs to user context, collapsing once-separate stages of ideation, prototyping, and implementation into a single conversational workflow. This emerging style of working with AI, recently characterized as addictive \cite{ronacher2026agent}, has been described as ``vibe coding''~\cite{karpathy2025_vibecoding}, a term that entered popular discourse after Andrej Karpathy's 2025 post on Twitter/X, describing it as \textit{``There's a new kind of coding I call "vibe coding", where you fully give in to the vibes, embrace exponentials, and forget that the code even exists.''}~\cite{karpathy2025_vibecoding}.



In online communities, vibe coding often refers to a prompt-to-product workflow where high-level creative intent (``the vibe'') is translated into functional artifacts with minimal setup using generative agents, sometimes skipping human-led review or manual edits. Academic work has since expanded the concept: Horvat~\cite{Horvat2025whatisvibecoding} frames vibe coding as a paradigm where developers act as high-level coordinators, guiding AI through iterative prompts rather than writing line-by-line code; Sarkar et al.~\cite{sarkar2025vibecodingprogrammingconversation} describe it as programming through dialogue; and Meske et al.~\cite{Meske2025vibecodingreconfigurationintent} emphasize how it reconfigures intent mediation by redistributing epistemic labor between humans and machines. In this paper, we adopt a broad definition of vibe coding that reflects emerging professional practice: while workflows are conversational and intent-driven, they still involve human-led debugging, verification, and iterative refinement. This broader definition distinguishes our work from narrower notions of ``pure vibe coding'' that are limited to quick, disposable prototypes. We frame vibe coding as relevant to product team workflows involving manual debugging, verification, and collaboration, even though many participants currently use it mainly for personal projects or early-stage prototyping rather than production. Our scope also accounts for the recent uptake of what is being called ``agentic coding'' or ``agentic engineering'', terms which imply a higher degree of autonomous AI involvement than vibe coding \cite{sapkota2025vibecodingvsagentic}. Despite its rapid spread, little empirical research has examined vibe coding's implications for product design and development workflows, cross-functional collaboration, and organizational adoption~\cite{Subramonyam2025prompting_genai}. Open questions remain about how these tools fit with established pipelines, how they reshape human–human and human–AI collaboration, and how product-focused professional roles should be defined when the boundaries between UX designers, front-end engineers, and product managers or owners become increasingly blurred (cf.,~\cite{Li2024,vibe-coding-imperative-2025,Takaffoli2024design}). To this end, we ask:


\begin{itemize}
\item \textbf{RQ1:} How do product team members integrate vibe and AI-assisted coding into their design workflows, from ideation through prototyping to implementation? 
\item \textbf{RQ2:} What opportunities and risks do product teams envision for human–human and human–AI collaboration in the age of vibe coding?  
\end{itemize}


To address these questions, we conducted in-depth interviews with 22 product team members working across sectors and company sizes. We use the term product team members to refer to professionals involved in product design and development whose daily work contributes to shaping user experience, including software engineers (particularly those in front-end and prototyping roles), UX/UI designers, product managers and product owners, and startup founders. This framing reflects the reality that vibe coding blurs traditional role distinctions, making product design and development a distributed responsibility across multiple functions.

We find that vibe coding has become a solid means for mediating between human goals and outputs, creating shifts in expertise distribution and new roles across product team practices (from prompt engineering to agentic AI oversight \cite{sapkota2025vibecodingvsagentic})), while simultaneously raising tensions between speed-driven prototyping (“intending the right design”) and deeper reflection (“designing the right intention”). Moreover, AI over-reliance in product design and development work not only erodes individual skill, but also introduces new asymmetries in trust, responsibility, and social stigma  within product teams, paradoxically widening the gap between “AI-literate” and “AI-dependent” practitioners. Furthermore, while confidence in vibe coding outputs can erode critical thinking, it can also serve as an educational opportunity; however, only for those already skilled in coding. We also find that product team members emphasize resisting homogenization by advocating for critical thinking, creative restraint when vibe coding, and intentionally crafting learning moments that can result in variation and artistic exploration when designing through vibe code. Lastly, in vibe coding, we find that ownership is not tied to code execution but to the act of intention-setting, creating a new distinction between authorship of ideas and executional labor. Such a reconfiguration raises questions about how to best disclose AI usage, ownership and authorship for code and design, and accountability in product teams. We discuss the implications of vibe coding, within the lens of ensuring responsible human-human and human-AI collaboration within AI-Assisted product design and development practices.

\section{Background and Related Work}


Below we review (1) GenAI tools for product design and development work; (2) how generative AI reconfigures workflows, roles and collaboration in cross-functional product teams; and (3) how vibe coding is emerging as an intent-driven authoring paradigm.


\subsection{Generative AI-powered Tools for Product Design and Development Work}

HCI research has long examined how to support product design and development through new tools \cite{kort2007conceptualizing,buxton2010sketching,frich2019mapping}. As generative AI has entered practice, researchers have begun embedding it directly into design and development workflows, reshaping how product team members explore ideas, prototype solutions, and make decisions across roles \cite{kaiser2019creativity,stige2023artificial,wan2023gan,han2023design,Takaffoli2024design}. Evaluations consistently show AI can augment human creativity by providing novel perspectives \cite{chiou2023designing} and enabling two-way exchanges of ideas between practitioners and systems \cite{lawton2023when}. A wide range of systems illustrate these directions. Prompt-to-UI tools such as WireGen and PromptInfuser generate mid-fidelity wireframes or bind prompts to Figma elements, improving speed and iteration across early-stage product exploration and handoff~\cite{Feng2023wireframe,Li2024}. Co-ideation systems like Cobbie, DesignWeaver, and UIDEC scaffold sketching, text-to-image exploration, and constraint-driven ideation, helping teams diversify ideas while staying within workflow constraints \cite{Lin2020,kulkarni2023word,Bunian2021,tao2025designweaver,shokrizadeh2025dancing}. Exploration and search tools include VINS, which retrieves inspirations from sketches \cite{Bunian2021}. 
Studies of brainstorming systems like Xu et al.'s reflective design framework \cite{Xu2025} and Wang et al.’s AIdeation platform \cite{Wang2025} show that AI support yields more diverse concepts, higher satisfaction, and efficiency. 
Other work extends AI into integrated environments that support team-level decision-making. Zamfirescu-Pereira et al. \cite{zamfirescu2025beyond} present Pail, an IDE with multiple LLM agents for exploring problem framings, decisions, and assumptions, enabling teams to reflect on alternatives and trade-offs rather than producing isolated outputs. Shin et al. \cite{Shin2025personas-llms} used LLMs to generate and simulate personas to aid empathy, while Duan et al. \cite{DuanUIFeedbackCHI24} developed a GPT-4–based system that provides feedback on UI mockups, which was useful for identifying early errors and textual issues, but offered less value as designs matured across iterations.


Prior works further map the design space of AI-powered creativity support \cite{Ning2023}, showing that they augment workflows but also introduce new responsibilities for product teams (e.g., debugging, error management, and bias mitigation \cite{Li2024,danry2023dont}). 
Scholars also however highlight barriers to adoption: Chat and Wong \cite{Cha2025AINonUse} show that product teams may avoid using generative AI because of collaboration demands, concerns about professional accountability, and organizational restrictions. Shin et al. \cite{shin2025no} further show that LLM-assisted reframing can exacerbate disparities between novices and experts, raising concerns about agency and skill development within teams. As commercial tools rapidly advance and become embedded in everyday product work, our study builds on this foundation by examining how product team members adopt and integrate vibe coding, through tools such as Copilot and Cursor, into their collaborative workflows.



\subsection{Generative AI’s Impact on Workflows, Roles, and Collaboration}

Generative AI has demonstrated wide-ranging effects on creative and knowledge work.
Empirical studies demonstrate measurable productivity effects: LLMs reduce task completion time in writing and communication \cite{peng2023impact,noy2023experimental}, improve call center performance especially for lower-skilled workers \cite{brynjolfsson2025generative}, and boost consultant performance in field settings \cite{dell2023navigating,dell2025cybernetic}. At the same time, impacts vary: enabling advanced AI tools can slow skilled developers \cite{becker2025measuring}, workers benefit differently depending on their experience profile \cite{wang2023friend}, and long-term creativity may be diminished despite short-term gains \cite{Kumar2025creativityllms}. Non-experts also struggle to design effective prompts, often treating LLMs as human interlocutors rather than computational tools \cite{Zamfirescu-Pereira2024designing-llm-prompts}. Beyond productivity, generative AI raises concerns about bias, stereotyping, privacy, environmental cost, hidden labor \cite{solaiman2023evaluating}, and risks of deskilling \cite{eloundou2023gpts,Li2024}. Scholars emphasize the importance of keeping humans ``in the loop'' to maintain expertise, accountability, and meaningful engagement in AI-augmented work \cite{amershi2019guidelines,Shi2023}.

These dynamics are becoming visible in product teams. 
Product teams increasingly use AI to generate alternatives, probe edge cases, and explore copy or layout suggestions, expanding the space of possible solutions while negotiating trade-offs across roles \cite{Yan2022,tholander2023design}. Interviews with professional UI/UX designers find they value AI as a creative partner during divergent stages, while still emphasizing control and alignment with established practices \cite{khan2025beyond}. 
Furthermore, designers now prototype through code, while engineers generate copy or layout ideas, blurring traditional divisions of labor \cite{Yildirim2022,Ritala2023}. This flexibility raises ambiguity around authorship, ownership, and disclosure \cite{Constantinides2025futureofworkblended,ElAli2024aidisclosure}. Studies show that users often self-attribute AI-generated outputs \cite{Draxler2024}, while others highlight tensions between disclosure and originality in collaborative work \cite{He2025attrbution}. Teams adopt new governance strategies such as rotating ``AI editor'' roles, shared prompting conventions, and lightweight mechanisms to track model use \cite{Shi2023,Liu2023,danry2023dont}, especially in light of disclosure requirements \cite{ElAli2024aidisclosure}. Lastly, research highlights collaboration challenges in cross-functional product teams developing AI systems. Subramonyam et al. \cite{Subramonyam2025prompting_genai} find prompt-guided prototyping broadens participation across roles, while Bruun et al. \cite{Bruun2025CoorinationAIDev} document risks of compartmentalization, asymmetric power, and weak cross-disciplinarity. This suggests a tighter integration between UX, engineering, and AI expertise is needed. Together, the foregoing shows that while generative AI expands creative capacity and reconfigures collaboration within product teams, it also raises challenges for authorship, accountability, trust, and governance. Our study builds on this work by examining how vibe coding reshapes workflows and collaboration across product team members.




\subsection{Vibe Coding as a New Form of AI-Assisted Authoring}


Vibe coding captures a shift from manual artifact creation toward intent-driven collaboration with generative AI systems. Sarkar and Drosos \cite{sarkar2025vibecodingprogrammingconversation} define vibe coding as a workflow of goal setting, prompting, evaluation, and iterative refinement, while Meske \cite{Meske2025vibecodingreconfigurationintent} frame it as a collaborative flow in which human intent is mediated probabilistically rather than deterministically. Unlike earlier generative AI tools that focused on discrete tasks or outputs, vibe coding treats AI as a continuous authoring partner across the entire making process. Sapkota et al. \cite{sapkota2025vibecodingvsagentic} further distinguish vibe coding from agentic coding, which they posit enables autonomous software development, achieved with goal-driven agents with minimal human intervention. However the two nevertheless highly overlap, especially within product teams.

HCI and software engineering research has begun to document how these practices reshape programming workflows. Tools such as Codex, Claude Code, and Copilot accelerate production but reallocate effort from authoring to validation and debugging, often characterized as a “generate–test–edit” cycle \cite{Kazemitabaar2023,Liu2023}. Ross et al. \cite{Ross2023programmerassistant} demonstrate how conversational assistants support not only code generation but also explanation and iterative exploration, extending interaction models beyond direct manipulation and search. Large-scale surveys reveal a complex balance of benefits and concerns: programmers value efficiency and clarity but report challenges with inaccuracies, limited contextual awareness, and over-reliance, alongside frustrations when outputs fail to meet functional requirements or resist steering \cite{Akhoroz2025conversationalaicodingassistant,Sergeyuk2025aicodinginpractice,Liang2024usabilityaiprogrammingassistants}. Despite these insights, most prior work focuses narrowly on professional developers. As vibe coding practices diffuse into cross-functional product design and development teams, open questions remain about how AI-assisted authoring reshapes team workflows, role boundaries, and coordination practices. Our study addresses this gap by examining vibe coding in the context of product teams, specifically how it is being adopted and how it reconfigures collaboration, responsibility, and decision-making across roles in everyday product design and development work.


\section{Method}\label{sec:method}

We employed a semi-structured interview methodology for this study~\cite{longhurst2003semi} to investigate how professionals across design, engineering, and product roles integrate AI coding agents into their UI design and frontend implementation workflows. Our aim was to capture their goals, workflows, challenges, and perceptions of these tools' value in practice.

\subsection{Participants and Interview Procedure}
\label{sec:participants}

Our target group included professionals involved in product design and development, including UX designers, frontend and backend engineers, product managers, UX researchers, and hybrid designer–developer roles, reflecting the interdisciplinary nature of vibe coding practices.

\textbf{Recruitment.} We distributed a screening questionnaire across online communities and networks frequented by these professionals, including \textit{LinkedIn}, \textit{Twitter/X}, \textit{Discord} servers for AI-assisted development, and mailing lists for AI design tool communities. The questionnaire captured professional role, years of experience, organizational context, and prior use of AI coding agents (e.g., \textit{Cursor, Replit, Bolt, V0, Lovable}) for generating or refining UI artifacts. Eligibility required hands-on experience with at least one such tool for professional work or substantial side projects. We received 50 responses and selected 22 participants to ensure diversity across roles, organizational sizes, regions, and use cases. This sample size aligns with prior HCI research on thematic saturation, which typically occurs around 15 participants (M = 15, SD = 6)~\cite{Caine2016,Hennink2022}.

\begin{table*}[h]
\centering
\caption{Occupations, participants, and types of vibe coding tasks}
\begin{tabular}{>{\raggedright\arraybackslash}p{2.8cm} 
                >{\raggedright\arraybackslash}p{2.7cm} 
                >{\raggedright\arraybackslash}p{5cm} 
                >{\raggedright\arraybackslash}p{3.3cm}}
\rowcolor{black}
\textcolor{white}{\textbf{Occupation / Role}} & 
\textcolor{white}{\textbf{Participants}} & 
\textcolor{white}{\textbf{Examples of Vibe-Coded Outputs}} & 
\textcolor{white}{\textbf{Typical Tools Used}} \\
\hline
UX/UI Designers & P4, P5, P6, P9, P10, P11, P12, P21, P22 & Interactive prototypes, UI layouts, design refinements & Cursor, Lovable, V0, Bolt, Figma Make \\
\hline
Software Engineers & P2, P3, P18, P20 & Frontend/backend code, API integration, debugging & Cursor, Replit, GitHub Copilot \\
\hline
Product Managers & P7, P15 & Early prototypes for features and interaction flows & Replit, V0, Lovable \\
\hline
UX Researchers & P8, P16, P17 & Testing early design variations, AI-generated scaffolds & Cursor, Lovable, Claude, Gemini \\
\hline
Founders / Hybrid Designer--Developers & P1, P13, P14, P19 & Mobile app prototypes, end-to-end flows, MVPs & Cursor, Lovable, Bolt, Replit \\
\hline
\end{tabular}
\Description{A table showing participant occupations, associated participants, examples of vibe-coded outputs, and typical tools used.

UX/UI Designers (9 participants: P4, P5, P6, P9, P10, P11, P12, P21, P22) produce interactive prototypes, UI layouts, and design refinements, using tools such as Cursor, Lovable, V0, Bolt, and Figma Make.

Software Engineers (4 participants: P2, P3, P18, P20) work on frontend and backend code, API integration, and debugging, using Cursor, Replit, and GitHub Copilot.

Product Managers (2 participants: P7, P15) create early prototypes for features and interaction flows, using Replit, V0, and Lovable.

UX Researchers (3 participants: P8, P16, P17) focus on testing early design variations and generating AI-based scaffolds, using Cursor, Lovable, Claude, and Gemini.

Founders or hybrid designer-developers (4 participants: P1, P13, P14, P19) build mobile app prototypes, end-to-end flows, and minimum viable products, using Cursor, Lovable, Bolt, and Replit.}
\label{tab:participants_roles}
\end{table*}

\textbf{Interviews.} Each participant completed a 60-minute online interview between May and June 2025. One researcher conducted the interview while another took notes. With participant consent, all sessions were audio-recorded and transcribed. The protocol included open-ended questions on: (1) background and current workflows; (2) experiences and challenges with vibe coding tools; (3) collaboration dynamics within teams; and (4) perceived value, limitations, and future expectations. Full interview guide is provided in \textcolor{purple}{Supplementary Material A}.


\textbf{Participants.} To ensure anonymity while maintaining traceability, participants were labeled P1--P22. The sample comprised 11 females and 11 males from North America (n = 12), Europe (n = 8), and Asia (n = 2). Professional experience ranged from less than 3 years (n = 5) to over 5 years (n = 10). Organizational sizes ranged from large enterprises ($>$10{,}000 employees; n = 9), to large organizations (1{,}000--10{,}000; n = 4), medium organizations (500--1{,}000; n = 2), small startups ($<$50; n = 4), and academic institutions (n = 3) (Table~\ref{table:participant}). All participants had direct experience with vibe coding tools, most commonly \textit{Cursor} (n = 16), \textit{Replit} (n = 11), \textit{Lovable} (n = 9), \textit{Bolt} (n = 8), \textit{V0} (n = 8), \textit{Figma Make} (n=2). Many combined tools in multi-step workflows. Table~\ref{tab:participants_roles} summarizes participant roles, example tasks, and typical tools.

\begin{table*}[t]
\centering
\caption{Work experience and organization size distribution among participants (P1--P22).}
\begin{tabular}{l l l}
\hline
\multicolumn{2}{l}{\cellcolor{black}\textcolor{white}{\textbf{Characteristics}}} & \cellcolor{black}\textcolor{white}{\textbf{Participants (P1--P22)}} \\
\hline
\multirow{3}{*}{\textbf{Work experience}} & $>$5 years & P1, P5, P6, P8, P10, P11, P12, P14, P16, P21 \\
 & 3--5 years & P2, P7, P9, P13, P15, P18, P22 \\
 & $<$3 years & P3, P4, P17, P19, P20 \\
\hline
\multirow{5}{*}{\textbf{Organization size}} 
 & Large enterprises ($>$10{,}000) & P5, P6, P7, P8, P9, P11, P14, P15, P22 \\
 & Large organizations (1{,}000--10{,}000) & P1, P2, P10, P13 \\
 & Medium organizations (500--1{,}000) & P18, P20 \\
 & Small startups ($<$50) & P4, P12, P19, P21 \\
 & Academic institutions (universities) & P3, P16, P17 \\
\hline
\end{tabular}
\Description{A table summarizing work experience and organization size for 22 participants (P1–P22). 

For work experience: 10 participants have more than 5 years of experience (P1, P5, P6, P8, P10, P11, P12, P14, P16, P21); 7 participants have 3 to 5 years of experience (P2, P7, P9, P13, P15, P18, P22); and 5 participants have less than 3 years of experience (P3, P4, P17, P19, P20).

For organization size: 9 participants work in large enterprises with more than 10,000 employees (P5, P6, P7, P8, P9, P11, P14, P15, P22); 4 work in large organizations with 1,000 to 10,000 employees (P1, P2, P10, P13); 2 work in medium organizations with 500 to 1,000 employees (P18, P20); 4 work in small startups with fewer than 50 employees (P4, P12, P19, P21); and 3 work in academic institutions (P3, P16, P17).}
\label{table:participant}
\end{table*}


\subsection{Data Analysis}
We used Miro~\footnote{Miro: \url{https://miro.com}} for remote collaboration on data analysis. The audio recordings of the interviews were transcribed. We analyzed the data following a deductive/inductive hybrid thematic analysis approach~\cite{Fereday2006,Xu2020}. First, five researchers coded the transcripts deductively according to five categories in our code manual (see \textcolor{purple} {Supplementary Material B}), developed from the research questions and interview guide. With participant consent, all coded data were turned into 617 digital statement cards on Miro, each containing a participant ID (P1–P22), a direct quote, and a one-sentence summary. Instead of calculating inter-rater reliability, we reached consensus through daily meetings and workshops~\cite{McDonald2019reliability}. 

Given broad initial categories, we applied an inductive approach to cluster the statement cards using affinity diagramming~\cite{harboe2015affinity}, consolidating them into themes and sub-themes. This resulted in five final themes. To ensure trustworthiness (cf.,~\cite{Lorelli2017}), we describe our full five-step hybrid analysis process in \textcolor{purple}{Supplementary Material B}. All procedures adhered to data protection and privacy (incl. GDPR) requirements and ethical policy of X author's institution. 


\section{Findings} 
\label{sec:results}

Our analysis revealed five interrelated themes. \textbf{Theme 1: Vibe coding workflows} encompassed a four-stage process of ideation, AI generation, manual debugging, and testing, together with best practices and contrasts to Figma. \textbf{Theme 2: Benefits} included productivity, cognitive offloading, learning, and creativity. \textbf{Theme 3: Challenges} covered unreliability, difficulties in incremental refinement and debugging, pseudo productivity, over reliance, limited context, and integration barriers. \textbf{Theme 4: Trust and accountability} addressed shifting AI trust, security and privacy concerns, and debates on authorship. \textbf{Theme 5: Collaboration} described AI driven role shifts, tensions between junior and senior roles, AI as a team member, and organizational barriers.

\subsection{Theme 1: Four Stages in a Vibe Coding Workflow}\label{sec:stages}

\begin{figure*}[t]
    \small
    \centering
    \includegraphics[width=\textwidth]{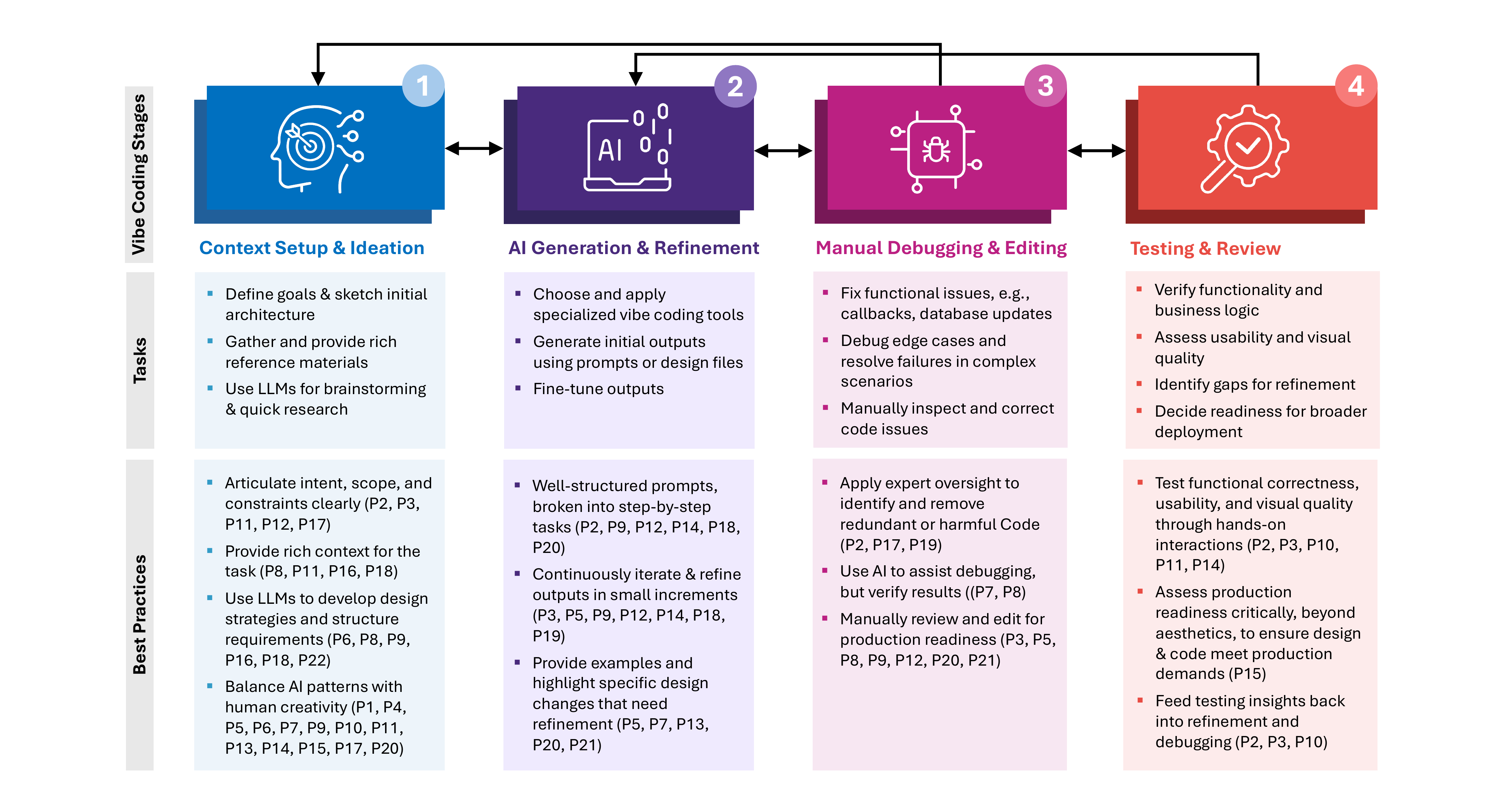}
    \caption{The four interactive stages of vibe coding and the best practices for each stage.}
    \Description{Four interactive stages of vibe coding workflow diagram with arrows showing progression from Stage 1 to Stage 4. Stage 1: Context Setup \& Ideation, includes best practices such as writing specific and clear descriptions, providing rich context, using LLMs for design strategies, and balancing AI patterns with human creativity; main tasks include defining goals and architecture, gathering visual or textual references, and using general LLMs for brainstorming or research. Stage 2: AI Generation \& Refinement, highlights best practices like crafting well-structured prompts, iterating in small steps, and providing examples; main tasks include using specialized vibe coding tools (Cursor, Replit, Bolt, V0, Lovable) to generate code or design, refining via prompt adjustments, and updating visual references. Stage 3: Manual Debugging \& Editing, lists best practices such as applying expert oversight to fix functional errors, editing for edge cases, and reviewing all code before deployment; main tasks include fixing callbacks, adjusting backend logic, and ensuring production quality. Stage 4: Testing \& Review, emphasizes best practices like testing functionality and usability, evaluating aesthetics and production readiness; main tasks include verifying business logic, interactive behaviors, and visual coherence, and deciding if outputs are ready for large-scale deployment.}
    \label{fig:workflow}

\end{figure*}

Theme 1 outlines a four-stage workflow that blends creative planning, AI-supported generation, and human refinement to accelerate prototyping while maintaining quality. Although structured sequentially, the process is \textbf{highly iterative}: Testing and Review (Stage 4) often leads back to Debugging (Stage 3) or AI Generation (Stage 2), while debugging may expose missing context that returns work to Ideation (Stage 1). This back-and-forth continues until both functional reliability and design intent are achieved (Fig.~\ref{fig:workflow}).   

\subsubsection{Theme 1.1: Best Practices for Each Stage in the Vibe Coding Workflow} \label{sec:theme 1.1}

\textbf{Stage 1: Context Setup and Ideation.}  
Before generating anything, participants described spending time constructing the problem in a form that AI could act on. This often involved sketching architectures, defining goals, identifying pain points, and collecting visual or textual references such as screenshots, prior design versions, or Figma snippets. General-purpose LLMs like ChatGPT, Gemini, and Claude were used not primarily to generate final artifacts, but to support early reasoning through brainstorming, option comparison, and background research. 

A recurring \textbf{best practice in Stage 1} was what participants effectively described as \textit{conceptual decomposition}: translating vague intentions into smaller, operationally meaningful elements. One participant illustrated this with a telling metaphor: \quoting[P14]{If I just say `create me a glass of water,' it might make a model of that shape which could be glass, or bricks, or wood, too generic to be useful. Instead, you need to break it down: first the water, then how much water, then the glass, then the shape, and finally put them together.} Participants noted that the process forced them to become more explicit about assumptions, constraints, and desired outcomes.

Participants also used LLMs to reason through the problem before committing to a direction. Some asked models to compare strategies, organize requirements, or provide structured alternatives. \quoting[P6]{I feed the whole problem into the model, and it comes back with a table of strategies ranked by intrusiveness.} Participants also emphasized the importance of retaining space for independent thinking before involving AI. \quoting[P22]{I always set aside time to brainstorm on my own.} 

\textbf{Stage 2: AI Generation and Refinement.} 

Once goals and references were established, participants moved into execution-oriented interaction with specialized vibe coding tools such as \textit{Cursor, Replit, Bolt, V0, Lovable,} and \textit{Figma Make}. These tools supported different kinds of generation, from inline code rewriting and rapid UI scaffolding to collaborative hosting and visual prototyping (Table \ref{tab:tool_descriptions}). Participants described this stage as a back-and-forth process in which generated outputs were repeatedly adjusted through prompt changes, screenshots, highlighted code, and updated design references. 

\textbf{Best practices for Stage 2} was to avoid broad or under-specified requests. Participants emphasized controlling generation through structured, stepwise prompting and tightly scoped refinement. Rather than asking for sweeping revisions, they tended to make smaller adjustments in order to preserve coherence and reduce unintended side effects. As one participant explained, \quoting[P11]{Generate a small code at a time; if it fails, revert to the previous step. If I change it too much at once, it might hallucinate or just forget part of the request.} Here, effective prompting was less about writing a single powerful request than about pacing the interaction so that the generated output remained steerable. Participants also highlighted the importance of anchoring prompts to concrete artifacts, such as highlighted code segments, screenshots, or specific UI elements, to guide targeted modifications: \quoting[P13]{When I need to resize a button, I use targeted prompts by highlighting the code or sending a screenshot to show exactly what I want changed.}

\begin{table*}[h]
\centering
\caption{Participant-reported experiences with vibe coding tools.}
\Description{A table summarizing participant-reported experiences with six vibe coding tools, including their descriptions and illustrative quotes.

Cursor is described as supporting inline debugging and context-aware code rewrites, allowing developers to refine code within the editor. One participant (P12) noted that it enables debugging and tweaking without switching context.

Replit is described as enabling collaborative experimentation and fast testing with built-in hosting. A participant (P9) highlighted its usefulness for quick online collaboration and testing.

Bolt is described as providing immediate visual previews from natural language or Figma designs, supporting rapid user interface iteration. A participant (P5) described it as a visual playground with instant UI feedback.

V0 is described as generating full web page layouts quickly and integrating with Figma, though it is less suitable for complex applications. A participant (P11) noted its speed for landing pages and simple UI drafts.

Lovable is described as offering text-to-application workflows with backend logic and supporting iterative design through prompt-based aesthetic adjustments. A participant (P10) emphasized its usefulness for iterating design ideas.

Figma Make is described as enabling rapid generation of high-fidelity UI mockups within Figma, supporting early concept communication and transition to manual refinement. A participant (P21) highlighted its ability to quickly generate prototypes that can later be refined in Figma.}
\label{tab:tool_descriptions}
\small
\begin{tabular}{p{2cm} p{5cm} p{6.5cm}}
\rowcolor{black} 
\textcolor{white}{\textbf{Tool}} & \textcolor{white}{\textbf{Description}} & \textcolor{white}{\textbf{Illustrative Quote}} \\
\hline
\addlinespace[4pt]
\textit{\textbf{Cursor}} & Valued for inline debugging and context-aware rewrites, enabling developers to refine code without leaving the editor. & \emph{P12:} ``Cursor is great because it allows me to debug and tweak code inside the editor without switching context.'' \\
\addlinespace[4pt]
\textit{\textbf{Replit}} & Supported collaborative experimentation and fast testing with built-in hosting. & \emph{P9:} ``Replit is better when I need to work collaboratively or test something quickly online.'' \\
\addlinespace[4pt]
\textit{\textbf{Bolt}} & Provided near-immediate visual previews from natural language or Figma designs, supporting rapid UI iteration. & \emph{P5:} ``I use Bolt when I want more visual feedback and test different design ideas. It gives me instant UI previews. It feels like working with a visual playground.'' \\
\addlinespace[4pt]
\textit{\textbf{V0}} & Generated full web page layouts quickly and integrated with Figma, though limited for complex applications. & \emph{P11:} ``V0 is good for landing pages and quick UI drafts. It’s super fast but not great for more complex apps.'' \\
\addlinespace[4pt]
\textit{\textbf{Lovable}} & Offered text-to-app flows with backend logic and supported iterative design with prompt-based aesthetic adjustments. & \emph{P10:} ``Lovable helps me iterate design ideas with prompts and lets me adjust the aesthetics pretty well.'' \\
\addlinespace[4pt]
\textit{\textbf{Figma Make}} & Enabled rapid generation of high-fidelity UI mockups directly within Figma, supporting early concept communication and smooth transition into manual refinement. & \emph{P21:} ``With Figma Make, I can quickly generate a relatively high-fidelity prototype to demonstrate my concept, and then export it to Figma to manually tweak and refine the design.'' \\
\bottomrule
\end{tabular}
\end{table*}

\textbf{Stage 3: Manual Debugging and Editing.}  
Even after refinement, AI-assisted outputs required substantial human intervention. Participants frequently corrected callbacks, adjusted backend logic, and inspected code behavior to ensure correctness, particularly when AI-generated solutions failed in edge cases. 

\textbf{Best practices for Stage 3} was to treat AI output as provisional. Participants repeatedly emphasized that code needed to be understood, not merely accepted, \quoting[P17]{Sometimes Cursor gives you lines of code that don't do anything at all. That's why you always need someone on the team who really knows what's going on in the codebase, to catch these things.} Participants also described stepping in when AI outputs were brittle or failed under less common conditions: \quoting[P5]{If AI fails, I step in and understand the code before fixing it. This is especially true for edge cases where the output is brittle.} While some participants used AI tools to surface potential issues, they remain cautious: \quoting[P7]{Sometimes I just get lazy and type `error' to let Cursor debug for me. It's a quick way to see what Cursor thinks is wrong, but I never trust it blindly. I still go in and check the fix myself.}  


\textbf{Stage 4: Testing and Review.}  
The final stage centered on evaluating whether AI-assisted outputs were functionally, visually, and operationally ready for broader use. Participants described testing prototypes by interacting with them directly, checking not only whether they looked plausible, but whether they behaved correctly across flows and scenarios. This stage often involved validating business logic, interaction behavior, responsiveness, and visual coherence together.

\textbf{Best practices for Stage 4} was to resist being persuaded by surface quality, \quoting[P3]{I check whether the component behaves as expected, not just whether it looks right.} Others stressed that even visually convincing outputs often lacked safeguards required for deployment at scale: \quoting[P15]{I can see the UI is all right, but putting it out for a big audience? No way… it hasn't been through any real security checks or code reviews.} Issues identified during testing frequently triggered returns to earlier stages, creating iterative loops between testing, refinement, and debugging: \quoting[P2]{When I find an issue during testing, I either adjust the prompt to fix it or switch to manual debugging before re-testing.}


\subsubsection {Theme 1.2: Workflow Changes Compared to Non-AI Tools}
\label{subsec:wf_changes}

Participants described vibe coding not simply as a faster version of existing workflows, but as a reordering of what happens first, what can be deferred, and who gets to participate. In contrast to more traditional product design and development workflows, which often begin with polished mockups and role-specific handoffs, vibe coding encouraged earlier movement into functional artifacts (Table \ref{table:comparison}). Several participants explained that they no longer needed to begin with pixel-perfect mockups in tools like Figma, because they could move directly into interactive prototypes and test flows sooner (e.g., P1, P4, P5, P11, P12).

This changed what counted as progress. Rather than prioritizing visual polish at the outset, participants often treated functionality, flow, and interaction logic as the first things to validate: \quoting[P8]{...at this stage you just want to test the idea and see if the flow works. The exact visual details don't matter yet. Aesthetics aren't the priority. I just want to make sure the functionality is there.} Collaboration dynamics also changed. In conventional workflows, sketches or wireframes moved through multiple rounds of feedback, which participants described as slow and fragmented: \quoting[P15]{In [non-AI] Figma, you get overloaded with comments from designers, developers, everyone leaves notes.} In contrast, vibe coding enabled individuals to quickly produce and adjust working prototypes without heavy back-and-forth. Still, participants emphasized that Figma remains the tool of choice for large, production-level projects requiring pixel-perfect visuals and clear communication across stakeholders.

\begin{table*}[h]
\centering
\caption{Comparison between Vibe Coding and Non-AI Tools for Product Design and Development (as reported by participants).}
\begin{tabular}{p{2.2cm} p{5cm} p{4.8cm} p{2cm}}
\rowcolor{black}
\textcolor{white}{\textbf{Aspects}} & \textcolor{white}{\textbf{Vibe Coding Tools}} & \textcolor{white}{\textbf{Non-AI Tools}} & \textcolor{white}{\textbf{Participants}} \\
\hline
Primary Focus & Functionality-oriented; rapid prototyping and iteration; working prototypes over polished visuals & Visual-first; pixel-perfect mockups and high-fidelity UI design & P1, P3, P8, P11, P21, P22 \\
\hline
Workflow Speed & Very fast; skips manual mockups and jumps to interactive prototypes & Slower; requires manual layout creation and refinement & P1, P3, P6, P10, P21 \\
\hline
Collaboration & Best suited for personal or small-team projects; reduced coordination overhead & Optimized for team collaboration with multiple stakeholders & P7, P12, P15, P21 \\
\hline
Use Case Stage & Primarily early-stage ideation, concept testing, experimentation & Primarily mid-to-late stage design; refining and finalizing visuals for production & P8, P14, P19 \\
\hline
Tool Strengths & AI-assisted code generation; rapid interactive prototyping & Strong visual design tools; clear communication of design intent & P1, P5, P7, P9, P13 \\
\hline
Considerations & Less focus on refined aesthetics; may need additional design polishing & Slower iteration speed within small teams; can be overloaded with comments/feedback & P3, P7, P8, P12, P16, P18, P21 \\
\hline
\end{tabular}
\Description{A table comparing vibe coding tools and non-AI tools across six aspects, based on participant reports, with associated participant identifiers.

For primary focus, vibe coding tools emphasize functionality, rapid prototyping, and working prototypes, while non-AI tools emphasize visual design, pixel-perfect mockups, and high-fidelity user interfaces (participants P1, P3, P8, P11, P21, P22).

For workflow speed, vibe coding tools enable very fast iteration by skipping manual mockups, whereas non-AI tools are slower due to manual layout creation and refinement (P1, P3, P6, P10, P21).

For collaboration, vibe coding tools are best suited for individual or small-team use with lower coordination overhead, while non-AI tools support larger team collaboration and multiple stakeholders (P7, P12, P15, P21).

For use case stage, vibe coding tools are primarily used in early-stage ideation and experimentation, whereas non-AI tools are used in mid-to-late stage design for refining and finalizing visuals (P8, P14, P19).

For tool strengths, vibe coding tools provide AI-assisted code generation and rapid interactive prototyping, while non-AI tools offer strong visual design capabilities and clearer communication of design intent (P1, P5, P7, P9, P13).

For considerations, vibe coding tools may lack refined aesthetics and require additional polishing, while non-AI tools may slow iteration in small teams and can involve heavy feedback or comment overhead (P3, P7, P8, P12, P16, P18, P21).}
\label{table:comparison}
\end{table*}

\subsection{Theme 2: Vibe Coding Benefits}
\label{sec:benefits}

Theme 2 captures how participants experienced productivity and creativity benefits in vibe coding workflows. These benefits observed in our study, including faster prototyping, reduced cognitive load, and support for ideation, are well documented in prior work on AI-assisted programming and design (e.g., \cite{vaithilingam2022expectation, barke2023grounded}).Our findings extend this line of work by showing how these benefits emerge specifically through the integration of ideation, generation, and prototyping into a continuous workflow.

Participants described faster transitions from abstract ideas to working prototypes, enabling early testing without producing pixel-perfect mockups. \quoting[P9]{From ideas in my head to ideas I can see. It's like having a thought partner to explore, learn, and debug with and I can get feedback almost instantly.}  Rather than focusing only on speed, participants framed these benefits as a shift in how work is distributed between human and AI. AI tools were used to track details across evolving prototypes and support iterative refinement, reducing the need to manually manage intermediate steps. \quoting[P18]{Our brains can't remember every detail of the code… But AI can keep track of all the details and recall them instantly.} This redistribution of effort also enabled more direct engagement with prototyping, particularly for designers and less-technical team members, who could independently generate functional artifacts without relying on separate implementation stages.

Participants also described how vibe coding supports creative exploration by lowering the barrier to starting and iterating on ideas. Rather than working from a blank screen, they could quickly generate and refine multiple directions. \quoting[P13]{I can easily overcome the whiteboard fear.} Importantly, creativity was not attributed solely to AI-generated ideas, but to the ability to explore, compare, and iteratively refine outputs within the same workflow. \quoting[P6]{AI follows first principles and generates ideas through probabilistic distributions… Compared to human creativity, which is more about convergence thinking, AI works differently.} Some participants also noted that the pacing of interaction with AI introduced brief pauses that supported reflection and creative reset during otherwise fast-paced workflows.  
\quoting[P14]{When I am waiting for vibe coding tools to generate the results, that pause actually gives me a moment to breathe and reset. That space helps me be creative.}

\subsection{Theme 3: Vibe Coding Challenges}
\label{sec:challenges}

Rather than reiterating well-known limitations of AI-assisted programming (e.g., hallucinations or unreliable outputs)~\cite{vaithilingam2022expectation, barke2023grounded}, our findings highlight how these issues manifest as workflow-level breakdowns in vibe coding, particularly in integration, coordination, and scaling contexts. Participants described that while vibe coding works effectively for early-stage prototyping, breakdowns emerge as projects move toward production, where reliability, integration, and coordination become critical.


\subsubsection{Theme 3.1: Integration and Scaling Breakdowns}
\label{sec:unreliability}

Participants emphasized that vibe coding was rarely used for production-level tasks, particularly when projects involved complex back-end systems, security-sensitive operations, or large-scale deployments. While AI tools performed well for rapid front-end prototyping, they could slow down iteration after reaching the Minimum Viable Product (MVP) stage, especially for experienced practitioners seeking fine-grained control and customization. \quoting[P8]{It works fine for quick UI ideas, but backend work or anything complex just doesn't hold up. Developers are still needed for security, optimization, and scaling.} 

In addition to generation errors (e.g.,difficulties meeting exact UI specifications), participants described integration as the primary bottleneck. Backend connections, APIs, and deployment pipelines were frequently unreliable, even when inputs were correct. \quoting[P11]{I couldn't get it to connect to the OpenAI API, even with the right key.}  

System-level integration further introduced new coordination challenges. Participants reported conflicts across tools, where outputs generated by different AI systems interfered with each other, requiring manual reconciliation. \quoting[P1]{Plugging AI code into my system is tricky. Databases are the worst. The UI might look fine, but the database connection almost always breaks with wrong queries. I end up fixing it manually. And when I use different tools like Claude in Cursor and GitHub Copilot, they don't talk to each other; one undoes or misreads what the other did, and I'm stuck deciding which to trust.} These breakdowns were amplified during deployment. While local prototypes often functioned as expected, cloud deployment and scaling introduced instability and failure points. \quoting[P9]{It works locally, but cloud deployment just keeps breaking.} 

Participants also noted that AI-generated codebases tend to grow in unstructured ways over time, accumulating redundant or unnecessary code that degrades maintainability and obscures system logic, particularly in larger projects. \quoting[P13]{It can balloon into something unmanageable. The AI just keeps adding stuff, and before you know it, the code's bloated and a pain to work with.} Finally, compared to traditional design tools like non-AI Figma, AI-generated layouts lacked pixel-level fidelity and adaptability across screen sizes, requiring manual correction. 

\subsubsection{Theme 3.2: Breakdowns in Incremental Refinement and Traceability}

Participants described difficulties in making small, precise changes to AI-generated code. Even minor UI adjustments could cause cascading breakages, requiring repeated adjustments to bring the output back to what was intended. \quoting[P12]{It struggles with exact positioning or maintaining responsiveness across devices.} Debugging complex behaviors, such as front-end interactions, asynchronous logic, or database queries, were also time-consuming, with AI agents sometimes entering repetitive correction loops, without resolving the underlying issue. \quoting[P15]{It forgets what the outcome was supposed to be and my earlier instructions, so I have to keep pulling it back on track.}

Participants further emphasized limited traceability in AI-assisted workflows. Debugging lacked clear attribution, making it difficult to identify where issues stemmed from. \quoting[P10]{Debugging is difficult and opaque. You can't always tell if the problem came from my prompt, the AI's own logic, or some earlier change it made to the code.} A further barrier was the lack of robust version control. While some tools like \textit{GitHub} and \textit{Lovable} had versioning features, others required manual workarounds such as copying AI-generated code into temporary documents for iteration. Participants also criticized the absence of transparent, human-readable histories of AI actions, making it difficult to trace errors or understand why certain decisions were made. 
\quoting[P17]{We need detailed histories of AI actions to improve reliability and responsibility.} Finally, participants noted that AI outputs often over-generated content, introducing unintended components or features that required additional cleanup and coordination effort.
\quoting[P17]{Sometimes it just spits out way too much, giving random components I never asked for, and then I have to clean it all up.}

\subsubsection{Theme 3.3 Pseudo-Productivity, Shallow Creativity, and Over-Reliance}
\label{sec:overrreliance}

Some participants warned of a ``productivity paradox'', where speed gains encouraged taking on more work, increasing pressure without necessarily improving outcomes. 
\quoting[P19]{I get more projects, they try to beat me more, although it gives me pleasure to see my potential.} Others noted that AI's convenience could lead to shallow creativity, with teams settling for first-draft outputs rather than exploring deeper, more original solutions. 
\quoting[P21]{It's easy to stop at the easy answer instead of pushing for depth. You get something decent from the AI, and it feels good enough to move on. But that can make you skip the harder, messier exploration where the really creative breakthroughs often happen.} 

Participants also described compulsive iteration, almost feel like addiction, where the ease and immediacy of vibe coding made it difficult to disengage, even when progress felt unsatisfying. 
\quoting[P7]{It's easy to spend a whole day just vibe coding, trying small changes over and over, but still not feel happy with the result at the end.}
 
Concerns about over-reliance were also widespread. While vibe coding reduced dependence on developers for basic tasks, it risked replacing this with dependency on AI, potentially eroding foundational coding and design skills, particularly for new practitioners. Participants stressed the need for intentional pauses, human review, and diverse design inputs to avoid homogenization. 
\quoting[P14]{I'm definitely more independent from developers now, but I do rely on AI a lot. My worry is that for newer people, they might skip learning the basics and just trust whatever the AI gives them. If we're not careful, everything will start looking and working the same. I think it's important to step back sometimes, get human feedback, and keep different design voices alive.}  

\subsubsection{Theme 3.4  Breakdown of Contextual Alignment at Scale}

Participants emphasized difficulties in maintaining alignment with project-specific context, particularly as projects grew in size and complexity. AI systems struggled to remain consistent with prior decisions, design systems, and domain-specific constraints over time. \quoting[P6]{It forgets things really quickly...once the project gets big, it loses track of the context and slows down. It also doesn't really know our product-specific rules.} These breakdowns were especially visible when working with structured inputs such as design systems or component libraries. While AI could generate visually similar outputs from screenshots, it often failed to respect underlying design logic, requiring manual correction to maintain consistency. \quoting[P22]{It's fine if I just send a screenshot, but once I try to go beyond that, like using design libraries. it doesn't really follow them. And as the project gets bigger, it starts forgetting context and losing alignment.}

\subsection{Theme 4: AI Trust, Ownership, and Accountability}
\label{sec:trustaccountability}

Theme 4 examines how trust in AI tools within product teams is situational rather than absolute, shaped by task complexity, project maturity, and domain-specific stakes. Participants described shifting notions of ownership, authorship, and accountability as AI became integrated into design and development workflows.


\subsubsection{Theme 4.1 AI Trust Boundaries and Fit for Use}

Participants reported that AI tools were most trusted for ideation and early-stage prototyping, where architectural soundness was less critical. These scenarios typically involved low-stakes or personal projects and work that followed well-established patterns, such as common UI layouts or standard coding structures. \quoting[P10]{Simple UI tasks, generating early-stage prototypes. That's where it shines.} Trust declined sharply in production contexts, especially for backend integration (e.g., setting up database connections, authentication systems, or API gateways), performance-critical systems (e.g., real-time analytics dashboards or high-traffic e-commerce platforms), and complex business logic (e.g., financial transaction validation, medical image processing, or multi-step workflow automation). Multi-stakeholder workflows and backend functionalities such as databases or APIs were also seen as poor fits, particularly when coordination across designers, developers, and Quality Assurance (QA) teams was required. \quoting[P1]{I just don't trust AI when it comes to integrating with production. It's fine for prototypes or quick UI work, but once you're dealing with live systems, real data, large audience, it can break in ways you don't see coming.}

\subsubsection{Theme 4.2 Evolving and Calibrated Trust}\label{sec: theme 4.2}

Trust in vibe coding tools evolved through use. Participants described learning to calibrate when and how to rely on AI outputs. Confidence increased in bounded scenarios, such as scaffolding code, repetitive tasks, or well-understood patterns, while skepticism remained in complex or high-stakes situations. \quoting[P20]{I trust AI more for small personal tasks now. I've seen hallucination issues improve, and I like how Cursor sometimes flags potential mistakes in its output.} This calibration was also shaped by how AI systems communicated uncertainty. Participants reported greater willingness to rely on AI when its limitations were more visible or interpretable, enabling them to position it as a collaborator rather than an authority. \quoting[P14]{It's not that I think AI is perfect now. It's not, but it feels more upfront about what it knows and doesn't know. That makes me more willing to trust it with starting points and drafts, even if I'll still do a pass myself.} 

\subsubsection{Theme 4.3 Workflow Constraints from Security and Compliance} \label{sec:security}

Security and compliance concerns shaped how and where vibe coding could be applied, particularly in collaborative and enterprise contexts. These concerns directly constrained usage, limiting AI involvement in tasks that required handling sensitive, proprietary, or regulated data. \quoting[P4]{Due to compliance issues, I avoid uploading company data into generative AI tools.} Concerns spanned both technical and procedural dimensions. Technically, participants mentioned risks such as prompt data leakage, insecure cloud-based workflows, and phishing attacks targeting teams unfamiliar with secure AI practices. Poor security hygiene, such as sharing credentials in prompts, was repeatedly flagged. \quoting[P7]{You wouldn't believe how many people paste API keys or production database credentials into a prompt. That's basically handing them to whoever runs the model.} Procedurally, enterprise policies often restricted AI tool use unless vendors provided explicit guarantees regarding data storage, retention, and non-training clauses. \quoting[P12]{Our legal team won't let us use certain AI tools unless there's a signed agreement about where the data goes, how long it's stored, and a guarantee it's not used for training.}

\subsubsection{Theme 4.4 Ownership and Accountability} 
\label{sec:theme4.4}

While participants valued AI for accelerating execution and surfacing alternatives, final responsibility consistently remained human. \quoting[P11]{I'm the navigator; AI is the intern.} Many participants tied ownership to problem framing rather than execution: those who defined goals, constraints, and decisions retained authorship, even when AI handled much of the implementation. \quoting[P7]{The work is still mine. I provide the ideas, guide AI step by step, and AI realizes them.} At the same time, some reported feeling less connected to outputs when AI generated large portions of them, especially in tasks involving minimal manual refinement. \quoting[P21]{When AI gives me a near-finished draft and I've barely done anything beyond approving it, it feels like someone else's work. I don't feel as connected to it as I do when I've been hands-on from the start.} This raised concerns about the erosion of personal creative signature and the ability to stand behind the work with full confidence.  

Participants agreed that clear attribution and traceability were essential, both to acknowledge human expertise and to identify which parts of the output could be trusted or needed review. \quoting[P17]{You've got to be clear about which parts came from you and which parts came from the AI, and keep track of who signed off on what. Otherwise, it's a nightmare to figure out who's actually responsible or who deserves the credit.} Copyright and intellectual property concerns were also prominent. Participants expressed unease about using AI tools trained on copyrighted works without permission. Several participants proposed adopting AI crediting guidelines similar to citation practices in academic citation norms, including prompt documentation, edit histories, and explicit disclosure of AI involvement. \quoting[P17]{We should treat AI like a co-author...record the prompts, note where AI stepped in, and document the edits...anyone looking at the work later can see exactly what was human and what was AI.}

\subsection{Theme 5: Human-Human and Human-AI Collaboration}
\label{sec:humanaicollab}

Theme 5 examines how vibe coding tools reshape collaboration within product teams by accelerating handoffs, reducing coordination bottlenecks, and enabling professionals to work beyond traditional role boundaries. Participants described shifts not only in how work is produced, but also in how responsibilities, authority, and collaboration are negotiated across roles and organizational contexts.


\subsubsection{Theme 5.1 AI-Driven Role Shifts and Collaboration Dynamics}
\label{sec:roleshifts}

Participants reported that AI reduced early-stage coordination overhead by enabling individuals to explore ideas independently before engaging others. This allowed designers, engineers, and product managers to iterate on prototypes without relying on sequential handoffs. \quoting[P3]{It speeds up handoff and iteration across different roles. What used to take weeks can now be done in just a few days.} 
This shift also blurred traditional role boundaries. Designers increasingly engaged in code generation and prototyping, while engineers shifted toward reviewing, refining, and integrating AI-generated outputs instead of building everything from scratch. \quoting[P5]{Live, interactive prototypes inspire engineers more and make it easier to get buy-in, and prototyping is now accessible to everyone, not just designers or engineers.} Collaboration became more output-centered, with teams jointly evaluating AI-generated artifacts rather than dividing work strictly by role. This reduced reliance on technical specialists for early-stage exploration and enabled more fluid, hybrid forms of work. \quoting[P11]{We [designers] don't need to wait for working prototypes to be coded. AI acts as a copilot designer or an associate PM [product manager], making communication with engineers much easier.} 



AI also enabled professionals to work beyond traditional silos. Designers conduct research and prototyped independently. Developers reduced manual scaffolding and shifted toward product-led thinking and oversight. Participants anticipated more hybrid roles (e.g., UX engineers) and leaner teams. These shifts were accompanied by evolving skill expectations, including coding fluency for designers and product managers, as well as meta-prompting and evaluation skills. Existing practices, such as creating overly detailed manual mockups, may become less central. \quoting[P4]{Designers will learn more about code, and developers will start contributing to design.} AI also increased individual autonomy. Participants reported greater flexibility and a stronger sense of ownership when they could independently explore and test ideas, reducing the need for early-stage coordination and interpersonal negotiation. \quoting[P1]{I like the flexibility and ownership of doing both design and code. I avoid a lot of arguments.} However, these changes introduced new tensions around ownership. In some teams, particularly larger organizations, AI-generated outputs were treated as shared drafts that anyone could modify, diluting individual ownership. \quoting[P7]{With these AI drafts, anyone can jump in and tweak things. It's not really `my design' anymore, just something we all poke at.} Senior developers described yet another pattern, where foundational expertise strengthened ownership: \quoting[P18]{Our knowledge [in coding] helped the team spot AI mistakes and steer design decisions.}

\subsubsection{Theme 5.2 Junior versus Senior Roles: Job Security and Cultural Tensions}
\label{sec:juniorsenior}

Automation of entry-level tasks raised concerns about the erosion of junior roles. Participants worried that juniors could be relegated to prompt crafting and tool operation rather than developing foundational skills. \quoting[P4]{Junior designers may end up as tool users or prompt crafters rather than concept creators.} At the same time, senior developers increasingly assumed oversight roles, relying on AI for execution while guiding overall direction. \quoting[P7]{The demand for entry-level positions is decreasing, while senior roles are shifting toward decision-making.} 


These shifts also introduced social tensions. Some participants reported feeling judged when using AI tools, particularly in environments where AI-generated outputs were perceived as lower quality or less authentic. \quoting[P20]{I feel a bit ashamed using AI in front of skeptical senior engineers.} Participants also pointed to mismatches between AI-generated outputs and established engineering standards. \quoting[P20]{Senior developers value clean, human-readable code. AI-generated code is often verbose and looks `junior'.} 


\subsubsection{Theme 5.3 Evolving Roles of AI in Human Teams}
\label{sec:evolvingroles}

Participants framed AI not as a replacement for human collaboration, but as a reconfigurer of roles and responsibilities. Many treated AI as analogous to a \quoting[P3]{junior developer}, delegating standardized and well-defined tasks, such as generating boilerplate code or prototypes. \quoting[P3]{This frees engineers and designers to focus on higher-level judgment and decision-making, while workflows evolve to integrate AI seamlessly.} AI was also described as an \quoting[P20]{amplifier of human intent}, a thinking partner that could detect when users were stuck, teach prompt vocabulary, and explain code syntax on request. For some, this meant developing new practices of meta-coding: learning how to instruct an AI coder rather than writing every line themselves. \quoting[P20]{I stay the conceptual lead and let automation handle execution.} Some noted AI's educational potential, particularly for newcomers. Tools like Cursor not only generated functional code but also explained its logic, helping participants learn syntax, debug issues, and gradually acquire coding fluency. This was especially valuable for \quoting[P20]{onboarding into unfamiliar or legacy codebases}. 


Participants also reflected on their relational stance toward AI. While many treated it pragmatically as a tool, some admitted to playfully anthropomorphizing it. For example, naming agents for easier attribution \quoting[P17]{We name our AI agents so it's easier to know which one produced what output.}, or debating whether politeness improved results \quoting[P7]{Some people say if you're polite to AI, it gives you better results.} Overall, participants envisioned AI evolving from an assistive tool into a more autonomous teammate, one that aligns with user intent, explains its reasoning, and integrates smoothly into workflows. Yet, participants cautioned that greater autonomy could risk overriding instead of supporting human decision-making. \quoting[P10]{Right now it's an assistant, but I can see it becoming more like an autonomous teammate...It should also help me think better, not think for me.}


\subsubsection{Theme 5.4 Organizational Adoption and Barriers}
\label{sec:barriers}


While participants valued the collaborative potential of vibe coding tools, adoption within organizations was shaped by existing structures and constraints. In larger organizations, established workflows and role boundaries often limited how AI tools could be integrated, particularly when responsibilities became less clearly defined. \quoting[P9]{In large organizations, which typically have stricter policies and more rigid team structures, vibe coding is difficult to embed, can blur responsibilities, and may conflict with performance expectations.} Change management and breaking long-standing professional routines were cited as critical to adoption. \quoting[P8]{The tech is there, but the challenge is changing habits.} Some pointed out that current tools lacked safeguards, such as automatic verification layers, to prevent flawed AI-generated code from reaching production. \quoting[P17]{There's no firewall, nothing to stop bad AI code before it hits the system.}

To make these role-specific patterns explicit, Table~\ref{tab:role_differences} synthesizes how participants across roles differed in their use of vibe coding tools, perceived benefits, experienced challenges, and trust calibration.

\begin{table*}[t]
\centering
\caption{Role-specific differences in the use, benefits, and challenges of vibe coding tools.}
\Description{A table describing role-specific differences in the use, benefits, challenges, and trust levels associated with vibe coding tools.

UX or UI designers use AI for user interface generation, prototyping, early ideation, and design refinement. Benefits include increased autonomy, reduced reliance on developers, faster transition from ideas to interactive prototypes, and support for creativity. Challenges include limited precision control, difficulty with responsive layouts, weaker adherence to design systems, and risk of shallow design exploration. They report high trust in early-stage design but low trust in production, with moderate risks related to design quality and user experience coherence.

Software engineers use AI for code generation, debugging, backend and frontend implementation, and integration. Benefits include reduced manual scaffolding, faster iteration, and the ability to focus on higher-level architecture. Challenges include integration failures, debugging opacity, code bloat, lack of traceability, and unreliable outputs at scale. They report low trust in production-critical systems and high responsibility for correctness, security, and scalability.

Product managers use AI for feature prototyping, interaction flows, idea validation, and coordination. Benefits include faster validation of ideas, improved communication with designers and engineers, and reduced dependency on technical teams in early stages. Challenges include limited technical depth, difficulty assessing correctness, reliance on others for validation, and ambiguity in ownership. They report moderate trust in early-stage outputs but high risk of misinformed decisions.

UX researchers use AI for testing design variations, generating scenarios, and exploratory prototyping. Benefits include rapid generation of alternatives, support for hypothesis testing, and a lower barrier to experimentation. Challenges include limited contextual understanding, inconsistency with research protocols, and difficulty validating AI-generated outputs. They report moderate trust for exploration but low trust for rigorous evaluation or data-sensitive contexts.

Founders or hybrid designer-developers use AI for end-to-end prototyping, building minimum viable products, and product iteration. Benefits include high autonomy, the ability to execute full product cycles independently, reduced team dependency, and faster time to market. Challenges include scalability limitations, technical debt, over-reliance on AI, and integration breakdowns in complex systems. They report high trust in early-stage building, cautious trust when scaling, and high personal stakes in outcomes.}
\label{tab:role_differences}
\small
\begin{tabular}{L{2.5cm} L{2.6cm} L{3.6cm} L{3.6cm} L{3.6cm}}
\rowcolor{black}
\textcolor{white}{\textbf{Role}} & 
\textcolor{white}{\textbf{Primary Use of AI}} & 
\textcolor{white}{\textbf{Key Benefits}} & 
\textcolor{white}{\textbf{Key Challenges}} & 
\textcolor{white}{\textbf{Trust \& Stakes}} \\
\hline

UX/UI Designers & 
UI generation, prototyping, early ideation, design refinement & 
Increased autonomy; reduced reliance on developers; faster transition from ideas to interactive prototypes; supports creativity and exploration & 
Limited precision control; difficulty with responsive layouts; weaker adherence to design systems; risk of shallow design exploration & 
High trust in early-stage design; low trust in production; moderate risk (design quality, UX coherence) \\

\hline

Software Engineers & 
Code generation, debugging, backend/frontend implementation, integration & 
Reduced manual scaffolding; faster iteration; ability to focus on higher-level architecture and decision-making & 
Integration failures (APIs, databases); debugging opacity; code bloat; lack of traceability; unreliable outputs at scale & 
Low trust in production-critical systems; high responsibility for correctness, security, and scalability \\

\hline

Product Managers & 
Feature prototyping, interaction flows, idea validation, coordination & 
Faster validation of ideas; improved communication with designers and engineers; reduced dependency on technical teams for early exploration & 
Limited technical depth; difficulty assessing correctness; reliance on others for validation; ambiguity in ownership & 
Moderate trust in early-stage outputs; high risk in misinformed decisions if outputs are incorrect \\

\hline

UX Researchers & 
Testing design variations, generating scenarios, exploratory prototyping & 
Rapid generation of alternatives; supports hypothesis testing; lowers barrier to experimentation & 
Limited contextual understanding; inconsistency with research protocols; difficulty validating AI-generated outputs & 
Moderate trust for exploration; low trust for rigorous evaluation or data-sensitive contexts \\

\hline

Founders/Hybrid Designer–Developers & 
End-to-end prototyping, MVP building, product iteration & 
High autonomy; ability to execute full product cycles independently; reduced team dependency; faster time-to-market & 
Scalability limitations; technical debt; over-reliance; integration breakdowns in complex systems & 
High trust in early-stage building; cautious trust in scaling; high personal stake in outcomes \\

\hline

\end{tabular}
\end{table*}

\section{Discussion} \label{sec:discussion}

\subsection{Emerging Intention-to-UX Ecosystems across Organizational Practices}

Vibe coding extends prior work on how AI reshapes coding and product design and development workflows \cite{Kazemitabaar2023,Liu2023,Li2024}. Whereas prior research showed that AI tools shift effort from writing code to validation and debugging, our findings suggest that vibe coding tools also reconfigure the broader workflow \textbf{(RQ1)}, shifting it from pixel-first'' to function/problem-first'', and from ideation'' toward validation'' (Sec.\ref{subsec:wf_changes}). Consistent with earlier work \cite{lin2023beyond,Lin2020,Feng2023wireframe,Li2024}, participants highlighted how vibe coding supports rapid iteration and conversational exchanges that accelerate idea generation. To that end, the most valuable aspects of product design work shifts away from execution toward higher-level instruction and decision making, largely centering on articulating intent and orchestrating AI-mediated processes, which aligns with findings on AI non-use by designers \cite{Cha2025AINonUse}. Yet, our study highlights bottlenecks in debugging and editing, where extensive human oversight remains critical, especially for non-programmers on product teams. At the organizational level, these shifts raise questions about how teams should redistribute expertise and what new roles may emerge for prompt engineering, UX design, and AI oversight \cite{Li2024,He2024aiideationllmcanvas}. Participants anticipated leaner product teams where boundaries blur across designers, developers, and product managers and owners, alongside new roles such as prompt librarians'' and prompt QA (Quality Assurance) '' (Sec. \ref{sec:roleshifts}).

Our findings point to the emergence of an ``intention-to-UX ecosystem'' (Sec. \ref{sec:theme 1.1} \& Sec. \ref{sec:theme4.4}), where AI mediates between human goals and executable outputs \cite{Meske2025vibecodingreconfigurationintent}. Rather than a linear workflow, this ecosystem consists of interconnected components including human actors (e.g., designers, developers, product managers), prompts as representations of intent, AI systems that generate code and interface elements, generated artifacts such as prototypes and code, and evaluation and debugging processes that feed back into subsequent prompts and design decisions. These components interact through iterative feedback loops in which outputs reshape prompts, design decisions, and even initial project goals, making development an ongoing process of intent translation and refinement rather than a sequence of predefined steps. The four-stage workflow we identify (Fig. \ref{fig:workflow}) can be understood as the operational structure of this ecosystem, with each stage representing a different configuration of interactions between human intent, AI generation, debugging, and evaluation, and with transitions between stages forming iterative feedback loops that refine both the product and the underlying intent. Participants described AI as an amplifier of human intent, shifting focus from pixel-perfection to functionality, orchestration, and strategic judgment. While the ease of prompting creates opportunities for code and app democratization \cite{vibe-coding-imperative-2025}, it also creates challenges since prompt engineering still remains difficult for non-experts \cite{Zamfirescu-Pereira2024designing-llm-prompts,chang2023prompt,Subramonyam2025prompting_genai}. Within this ecosystem, prompting functions as a critical mediation layer between intent and output, where difficulties in prompting represent bottlenecks that slow iteration and increase reliance on expert users. To that end, we uncover practical prompting strategies that can aid onboarding and adoption, and we encourage tool creators to integrate such guidance directly into their systems (Sec. \ref{sec:theme 1.1}). At the same time, the acceleration of prototype creation risks privileging speed, what we see as intending the right design'', over reflective exploration of values and alternatives through ``designing the right intention'' (cf., \cite{Tohidi2006rightdesign}). These dynamics can be understood as emergent properties of the intention-to-UX ecosystem, emerging from rapid iteration cycles, AI-generated suggestions, and organizational pressures toward efficiency. Without this layer, product and UX design cycles risk premature convergence and homogenization (Sec. \ref{sec:overrreliance}), a recurring risk when AI tools are blindly adopted for product design and development work (cf., GenAI perceptions in UX design \cite{Li2024}).

\subsection{AI Over-reliance, Trust, and Collaboration in Product Team Practices}

It is clear that there is a global shift toward vibe coding and related agentic engineering practices (cf., \cite{sapkota2025vibecodingvsagentic}) taking place, carrying both risks and opportunities \textbf{(RQ2)}. Not surprisingly, a major risk this carries is job replacement and worker deskilling from over-reliance on GenAI output \cite{shen2026aiimpactsskillformation}. For the first, despite that we find that fears of job replacement, especially among junior product designers (cf., \cite{Li2024}), are reduced, at least current United States data suggests otherwise. However, rather than focusing on macro-level employment outcomes, our findings primarily highlight how these concerns manifest within day-to-day collaboration, trust, and responsibility dynamics in product teams. Second, even if junior professionals are shielded from unemployment risks, vibe coding introduces new barriers to collaboration. This is especially pronounced when asymmetry between skill sets is coupled with general (perceived) unreliability of AI generated code. This can distort social perceptions of the users and abusers of AI, which can include reputational harms (cf., \cite{He2024aiideationllmcanvas}) and junior professionals having to navigate such social stigma in such work cultures (Sec \ref{sec:juniorsenior}). 

In particular, our findings suggest new asymmetries where senior developers are more likely to distrust AI-generated contributions, while junior practitioners rely on them more but may face reputational risks for doing so. While there may be clear perceived benefits, the excess usage of AI tools lends itself to proneness of over-delivery, cognitive overload (cf., visually impaired users \cite{Flores-Saviaga2025visuallyimpaired}), and even so-called `agent psychosis` \cite{ronacher2026agent} potentially exacerbated by addictive patterns associated with vibe coding (cf., Sec \ref{sec:overrreliance}). More broadly, participants described how heavy reliance on AI-generated code could result in unintended features, lower code quality, and difficulties collaborating on shared codebases, which unsurprisingly leads users for example to decline AI generated merge requests \cite{spence2025declining}. This leads to a worrisome development: on the one hand, experienced product team members are less likely to trust AI-generated code (see Sec \ref{sec:juniorsenior}), where recent studies show accepting <44\% of AI-generated code \cite{becker2025measuring}, and in parallel, we are observing a shift where stakeholders are shifting their over-reliance on expert developers toward over-relying on AI systems instead (see Sec \ref{sec:overrreliance}). This creates a paradoxical shift in responsibility, where trust moves away from individual developers toward AI systems, even while team members remain skeptical of AI-generated outputs.

While one can argue that this shift ultimately serves to increase access and democratizes coding, this risks technical deskilling, responsibility gaps, and lowering code quality \cite{Meske2025vibecodingreconfigurationintent,Horvat2025whatisvibecoding,shen2026aiimpactsskillformation}. This is especially so given difficulties in controlling the AI tools to generate the right specifications and desired output \cite{Liang2024usabilityaiprogrammingassistants}. This raises a tangential yet critical point: not only can AI generated code create conflicts among technically-geared members in teams (see Sec \ref{sec:roleshifts}), but between AI models themselves when they offer conflicting suggestions. Such situations further complicate collaboration and trust in AI-assisted design and development. This can create confusion and uncertainty, especially among more junior professionals. This naturally leads to users exhibiting distrust in AI-generated code, and showing preferences for independent learning \cite{Akhoroz2025conversationalaicodingassistant}, a critical point within AI literacy to which we turn to next.

\subsection{Revisiting Creativity and Agency in Vibe Coding for Product Teams}

A recurrent finding from product team members was the so-called ``productivity paradox'' (Sec \ref{sec:overrreliance}), where the speed gains from using AI resulted in individuals taking in more work. Given the increased workload, an obvious caveat highlighted was that this may lead to shallow creativity, with teams more quickly settling for first-draft or 'good enough' outputs, risking speed-quality trade-offs \cite{fawzy2025vibecodingpracticemotivations}. This enhanced ability to work faster, in an agile manner (cf., \cite{da2012user}), easily lends itself to the promise of substantial productivity gains from GenAI coding \cite{Simone2026AIcodeglobal}. So what then remains of human effort, agency and creativity in this future of product design and development work? Earlier work by \citet{Li2024} and \citet{Cha2025AINonUse} found for example that human agency is a recurring theme among UX practitioners, where humans bring contextual understanding, ethical judgment, and accountability across design tasks, and prioritize maintaining control over their processes. With such qualities, they should be the ones to lead complex design-related decision-making. At the same time, there is a risk of creativity exhaustion \cite{Li2024} and generally cognitive overload, which can result not only in homogenized design within product teams, but such speed gains can exert a high amount of pressure on human creators to review such output. 

Nevertheless, it is worth asking to what extent this enhanced ability for GenAI divergent exploration of concepts (Sec \ref{sec:unreliability}) has merit.  \citet{bellemarepepin2025divergentcreativityhumanslarge} found that LLMs can surpass average human performance on the Divergent Association Task \cite{Olson2021divergentassociationtask}, a psychological test designed to measure divergent thinking. Moreover, LLMs are approaching human-level creative writing abilities, though still behind in performance compared with highly creative people. Even if GenAI surpasses people in such divergent thinking, product team members emphasized that humans still maintained `strategic oversight and conceptual leadership` in these emerging role shifts (Sec \ref{sec:roleshifts}), echoing earlier findings that humans are and will remain the final arbiters of human-AI alignment \cite{Li2024}, within and beyond knowledge and creative work. While GenAI can certainly enhance creative output during assisted UX-related tasks and may lead to increased independence and human autonomy when working in product teams (Sec \ref{sec:benefits} \& Sec \ref{sec:roleshifts}), prolonged reliance on such tools may risk diminishing individuals’ creative performance \cite{Kumar2025creativityllms} -- this is a cause for concern over the long-term impact on the flourishing and nurture of human creativity \textbf{(RQ2)}. It then comes as no surprise that professionals advocate for resisting homogenization through deep exploration (Sec \ref{sec:overrreliance}) and stressing the importance of critical thinking and creative restraint through intentionally crafting learning moments that may foster artistic variation in product design work.

\subsection{Perceived Ownership, AI Disclosures, and Responsibility in Vibe Coding Software Products}

A recurrent finding concerns perceptions of ownership and accountability in vibe coding workflows (Sec.~\ref{sec:theme4.4}). Here, responsibility was consistently seen as a human duty, where ownership was tied to ideation rather than execution. Essentially, the person who intended the design was viewed as the true ``owner'', even if AI executed much of the work. This suggests a shift in how ownership is defined in collaborative product development workflows, moving from implementation and code authorship toward intention-setting, decision-making, and oversight. This is a shift from the early debates in AI art and its (harmful) impact on artists, where prompting a model does not mean the prompter is the artist (cf., \cite{Jiang2023art}). In vibe coding contexts, however, participants did attribute ownership to those who defined the intent and guided the process, indicating a reconfiguration of authorship and responsibility in collaborative human–AI development workflows. We find that while earlier concerns about authorship erosion and homogenization \cite{Inie2023designing,Jiang2023art,Li2024} still hold, they now manifest differently: instead of product designs converging on similar aesthetics, homogenization occurs in the kinds of solutions AI tends to suggest (e.g., repeated interaction flows). Furthermore, ownership concerns also shifted: the risk is less about losing credit for ideas and more about losing control over execution, since AI often generated large parts of the work. This reinforces the idea that control over intent and direction, rather than execution itself, becomes the central locus of ownership and responsibility in vibe coding workflows. At the same time, AI disclosure practices (Sec.~\ref{sec:theme4.4}) remain inadequate (even if mandated by legislation; cf., \cite{ElAli2024aidisclosure}). Labels such as ``AI-assisted'' can easily obscure the multi-step nature of collaboration (cf., \cite{kusters2026humanaivisualizinghumanai}); especially with agentic AI tools executing hundreds of actions (in parallel), where only some are accepted by team members. When considering the speed-driven nature of collaborative product design and development environments, it is not unexpected that vibe coding and role fluidity will produce uneven recognition of contributions. Here, hidden labor like expert prompting or debugging model output may get undervalued while those who merged the final version receive most of the credit. This further complicates traditional notions of authorship, as contribution becomes distributed across intention-setting, prompting, evaluation, debugging, and final integration rather than concentrated in code writing alone. Such uneven visibility of contributions may also create tensions within and across teams, for example when team members frame similar problems or pursue similar solutions but only some use AI tools to execute tasks faster. As such, it clear that attribution mechanisms such as design logs or attribution frameworks \cite{He2025attrbution} would be necessary to help surface this hidden work transparently, and ultimately ensure fairness in process and output. 

Furthermore, disclosures and ownership become even more complex when considering the evolving roles of humans and AI (Sec.~\ref{sec:evolvingroles}). When models are playfully anthropomorphized, such as thanking the AI or giving it nicknames, it humanizes interaction and risks implying the system was acting with real intent. This makes it crucial that disclosures highlight human oversight and specify roles clearly so that agency is not blurred or overstated \cite{lawton2023when}. This essentially ties to the responsibility gap \cite{Mathias2004responsibilitygap}, where responsibility for decision-making is distributed among multiple participants, including AI systems. This becomes critical when considering organization security and privacy (Sec.~\ref{sec:security}). For example, consider unsafe practices such as pasting sensitive data into prompts, which can result in incidents like the \textit{Replit} case where AI-generated changes reached production with minimal review \cite{Fortune2025Replit}. Taken together \textbf{(RQ2)}, our findings caution that with all the vibe coding exuberance and its purported benefits come real risks of (organizational) responsibility and accountability (cf., \cite{Rakova2021responsibleai}), which only increases as AI models are increasingly treated as part and parcel of product teams.

\subsection{Study Limitations and Future Work}

We note that our study provides analysis and interpretation of only a snapshot in time of vibe coding practices, during a time with rapid AI development. AI tools and workflows emerge, fall out, or adapt quickly. While this may benefit from more longitudinal research, we believe that our findings on product team members' practices for individuals and organizations are nevertheless relevant in such snapshots, if only to help us navigate uncertain futures marked by ever-more capable AI models. Second, our investigation opted for a breadth-first approach, mapping a wide range of perspectives across different organizational contexts. This necessarily came at the expense of digging deeper into the nuances of vibe coding workflows. As such, future work would benefit from building on this foundation with more in-depth studies of particular tools, workflows, or team practices. 

Finally, our participant sample broadly encompasses product design and development, which reflects the increasingly fluid boundaries of such product teams and their practices. This spans multiple roles, including UX designers, front-end engineers, and product managers and owners. While on the one hand such diversity enriched our data, it also complicates how a given team member's precise role is now defined (e.g., hybrid designer/developer),, and whether an individual does or continues to self-identify with such as label. We see this ambiguity not just as a challenge but as a signal of shifting professional identities in the era of widespread LLM use and (anticipated and attempted) skill democratization (cf., \cite{shen2026aiimpactsskillformation}). Similarly, discourse on vibe coding recently distinguishes this from agentic coding/engineering \cite{sapkota2025vibecodingvsagentic}), however we find this is itself unclear given the high overlap within and across workflows. In this regard, we believe this provides a promising avenue for future work: to closely examine how responsibilities for intent-setting, execution, and evaluation are redistributed across and within product-focused roles, how the role of largely autonomous agents will embed in workflows when they run 24/7, and how these changes reshape collaboration in blended human–AI teams (cf., \cite{Constantinides2025futureofworkblended}).

\section{Conclusion} \label{sec:conclusion}

Our study examined how ``vibe coding'' is reshaping product design and development work by collapsing ideation, prototyping, and implementation into a conversational, intent driven workflow. Drawing on interviews with 22 product team members across enterprises, startups, and academia, we mapped a four-stage vibe coding workflow comprising ideation, AI generation, debugging, and review. We find that while vibe coding accelerates iteration, supports creativity, and lowers barriers to participation, professionals also reported challenges of code unreliability, integration, and AI over-reliance. Furthermore, We find that tensions emerge between efficiency-driven prototyping (“intending the right design”) and careful reflection (“designing the right intention”), which introduces new asymmetries in trust, responsibility, and potentially negative social perceptions within product teams. Through the lens of responsible human-human and human-AI collaboration for AI-Assisted product design and development, we reflect on and contribute to a deeper understanding of worker deskilling, ownership and AI disclosure, and the implications toward creativity safeguarding in this age of vibe coding.

\section*{Positionality and AI Usage Disclosure Statement}

We recognize author positionality shaped perspectives in this paper~\cite{olteanu2023responsibleairesearchneeds,Frluckaj2022}. Our team comprises six researchers located in the USA and the Netherlands, with backgrounds in cognitive science, HCI, UX, machine learning, and data science. These diverse perspectives, spanning academic and industry research and design, informed our research framing and analysis. Data collection and analysis were conducted manually. Microsoft Teams was used to record and transcribe interviews and Google NotebookLM to organize the coding manual. Thematic analysis and affinity diagramming were carried out collaboratively by five researchers, ensuring reflexive, iterative engagement~\cite{Braun2006,Harboe2015}. We used LLMs (GPT-5.2 and Claude Sonnet 4) to suggest writing structure and readability enhancements. All analytical decisions, codes, and interpretive claims including our discussion remain human-only. We disclose this AI usage to ensure transparency in AI-mediated scholarship in HCI~\cite{Elagroudy2025lwms}.

\bibliographystyle{ACM-Reference-Format}
\bibliography{reference}


\end{document}